\documentclass[iop]{emulateapj}
\usepackage{apjfonts}
\usepackage{times,graphicx,aas_macros}
\usepackage{amsmath}
\usepackage{float}
\usepackage{ifthen}
\usepackage{times}
\usepackage{natbib}
\usepackage{rotating}

\begin{document}

\shorttitle{The {\it Herschel} Filament in RCS\,2319+00}
\shortauthors{Coppin et al.}

\title{The {\it Herschel} Filament: a signature of the environmental drivers of galaxy
evolution during the assembly of massive clusters at {\it Z}=0.9\footnote{{\it Herschel} is an ESA space observatory with science instruments provided by European-led Principal Investigator consortia and with important participation from NASA.}}

\author{K.~E.~K.~Coppin\altaffilmark{1}}
\author{J.~E.~Geach\altaffilmark{1,2}}
\author{T.~M.~A.~Webb\altaffilmark{1}}
\author{A.~Faloon\altaffilmark{1}}
\author{R.~Yan\altaffilmark{3}}
\author{D.~O'Donnell\altaffilmark{1}}
\author{N.~Ouellette\altaffilmark{4}}
\author{E.~Egami\altaffilmark{5}}
\author{E.~Ellingson\altaffilmark{6}}
\author{D.~Gilbank\altaffilmark{7}}
\author{A.~Hicks\altaffilmark{8}}
\author{L.F.~Barrientos\altaffilmark{9}}
\author{H.K.C.~Yee\altaffilmark{10}}
\author{M.~Gladders\altaffilmark{11}}

\altaffiltext{1}{Department of Physics, McGill University,
3600 Rue University, Montr\'eal, QC, H3A 2T8, Canada}
\altaffiltext{2}{Banting Fellow}
\altaffiltext{3}{Center for Cosmology and Particle Physics, Department
  of Physics, New York University, 4 Washington Place, New York, NY,
  10003, USA}
\altaffiltext{4}{Department of Physics, Engineering Physics \&
  Astronomy, Queen's University, Kingston, ON, K7L 3N6, Canada}
\altaffiltext{5}{Steward Observatory, University of Arizona, 933 N. Cherry Ave, Tucson, AZ 85721, USA}
\altaffiltext{6}{Center for Astrophysics and Space Astronomy, Department of Astrophysical and Planetary Sciences, UCB-389, University of Colorado, Boulder, CO, 80309, USA}
\altaffiltext{7}{South African Astronomical Observatory, P.O. Box 9, Observatory, 7935, South Africa}
\altaffiltext{8}{Department of Physics and Astronomy, 3255 Biomedical
  and Physical Sciences Bldg., Michigan State University, East
  Lansing, MI, 48824-2320, USA}
\altaffiltext{9}{Departamento de Astronom\'{i}a y Astrof\'{i}sica
Pontificia Universidad Cat\'{o}lica de Chile, Vicu\~{n}a Mackenna 4860, 7820436 Macul, Santiago, Chile}
\altaffiltext{10}{Department of Astronomy \& Astrophysics, University
  of Toronto, 50 St. George St., Toronto, ON, M5S 3H4, Canada}
\altaffiltext{11}{Department of Astronomy \& Astrophysics, University of Chicago, 5640 S. Ellis Ave, Chicago, IL, 60637, USA}

\begin{abstract}
We have discovered a 2.5\,Mpc (projected) long filament of
infrared-bright galaxies
connecting two of the three $\sim5\times10^{14}$\,M$_\odot$ clusters
making up the RCS\,2319+00 supercluster at $z=0.9$. The filament is revealed
in a deep {\it Herschel} Spectral and Photometric Imaging REceiver (SPIRE) map
that shows 250--500\,$\mu$m emission associated with a spectroscopically
identified filament of galaxies spanning two X-ray bright cluster cores. We
estimate that the total (8--1000\,$\mu$m) infrared luminosity of the filament is
L$_{\rm IR}\simeq5\times10^{12}$\,L$_\odot$, which, if due to star formation alone,
corresponds to a total SFR$\simeq900$\,M$_\odot$\,yr$^{-1}$. We are witnessing the
scene of the build-up of a $>10^{15}$\,M$_\odot$ cluster of galaxies, seen
prior to the merging of three massive components, each of which already
contains a population of red, passive galaxies that formed at $z>2$. The
infrared filament demonstrates that significant stellar mass assembly is taking
place in the moderate density, dynamically active circumcluster environments
of the most massive clusters at high-redshift, and this activity is concomitant
with the hierarchical build-up of large scale structure. 
\end{abstract}


\section{Introduction}

The average rate of galaxy growth, measured by the volume averaged star
formation rate (SFR), was a factor of $\sim$10 higher at $z\sim1$ than it is
today (e.g.,~\citealt{Hopkins06}), when the environments destined to become
the most massive ($10^{15}$M$_\odot$) clusters were still in the process of
assembly. What effect does the growth of large scale structure at $z\sim1$
have on the star formation histories of galaxies bound to, or being accreted
onto, such environments? For over 30 years it has been known that the fraction
of blue star-forming galaxies in clusters was higher in the past
(\citealt{BO78,BO84}), and more recent work has extended star formation
surveys of distant clusters into the infrared (IR) regime (important for
tracking the total SFR), reinforcing the view that there has been strong
evolution in the total SFR of clusters since $z\sim1$
(e.g.,~\citealt{Geach06}; \citealt{Bai09}). To some extent the evolution seen
in cluster populations tracks the field, but it is important to consider the
environmental context of this evolution (e.g.,~\citealt{GilBal08}), since
star-forming galaxies within high-{\it z} clusters are destined to evolve into
the `red sequence' in the cores of the descendants of such environments today
(e.g.,~\citealt{Gilbank08LF}; \citealt{Poggianti08}).

It is becoming clear that clusters exist at $z\sim1$ that already contain an
established population of massive galaxies that formed their stellar
populations quickly at much higher redshifts ($z>2$) when the cluster was in a
much earlier state of collapse (e.g.,~\citealt{Papovich10}). While this rapid
formation episode puts in place the massive tail of the cluster galaxy
population, the dwarf end of the red sequence undergoes significant evolution
via the continuous accretion of satellite galaxies, with the bulk of such
assimilation in the form of dwarf galaxies at $z=0$, steepening the faint-end
slope of the luminosity function (e.g.,~\citealt{DeLucia04};
\citealt{Gilbank08LF}). What is the intermediate stage of red-sequence
evolution at $z\sim1$ and what role, if any, does the {\it assembly} of the
clusters themselves play on their member galaxies? For example, is star
formation triggered by, or at least associated with, the peripheral
environments (outlying groups and filaments) feeding those massive clusters?

In this Letter we discuss observations performed with the ESA {\it Herschel
Space Observatory} \citep{Pilbratt10} Spectral and Photometric Imaging
REceiver (SPIRE; \citealt{Griffin10}). We have discovered a
remarkable filament of far-IR bright galaxies linking two of three massive
clusters in close proximity at $z=0.9$ -- the RCS\,2319+00 supercluster
\citep{Gilbank08}. Our study reveals a glimpse of the process of galaxy
evolution that is coeval with the birth of one of the most massive structures
in the Universe; a text-book example of structure assembly in the hierarchical
paradigm. We assume cosmological parameters of $\Omega_\Lambda=0.73$,
$\Omega_\mathrm{m}=0.27$, and $H_\mathrm{0}=71$\,km\,s$^{-1}$\,Mpc$^{-1}$
\citep{Spergel03}.

\section{RCS\,2319+00}

RCS\,231953+0038.0 (hereafter RCS\,2319+00) was detected in the first
Red-Sequence Cluster Survey (RCS-1; \citealt{GladdersYee05}). Subsequent
follow-up observations with {\it Chandra} \citep{Hicks08} and an extensive
optical and near-IR spectroscopic campaign (\citealt{Gilbank08}; A. Faloon
et al., 2012 in preparation; R. Yan et al., 2012 in preparation) have
revealed a remarkable supercluster system (RCS\,231953+0038.0, RCS\,232002+0033.4, RCS\,231948+0030.1) comprising three
distinct X-ray luminous cores with
$L_X\sim3.6$--$7.6\times10^{44}$\,erg\,s$^{-1}$, separated by $<3$\,Mpc in the
plane of the sky and $\sim10$\,Mpc along the line of sight (assuming a Hubble
flow). X-ray, strong lensing and virial mass estimates all imply individual
cluster masses of $\sim5\times10^{14}$\,M$_\odot$. The close
proximity of the three components, taken with the fact that the extended X-ray
emission and density profiles of the red-sequence members appear to be aligned
strongly suggests that the clusters are most likely in the early stages of a
three-way merger that will result in a $>10^{15}$\,M$_\odot$ cluster by
$z=0.5$ \citep{Gilbank08}.

\begin{figure*}
\centerline{\includegraphics[width=0.475\textwidth]{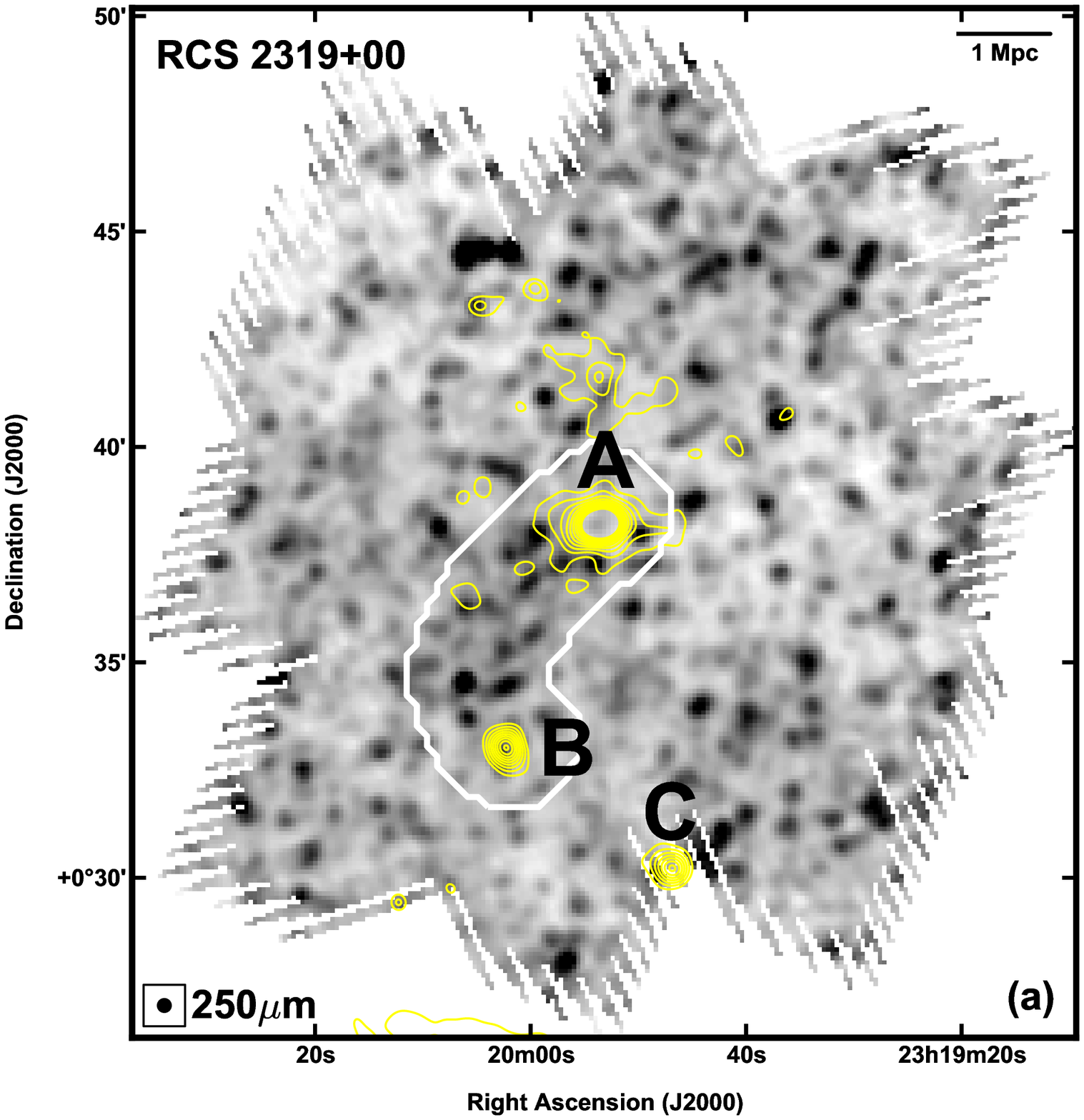}\hfill\includegraphics[width=0.475\textwidth]{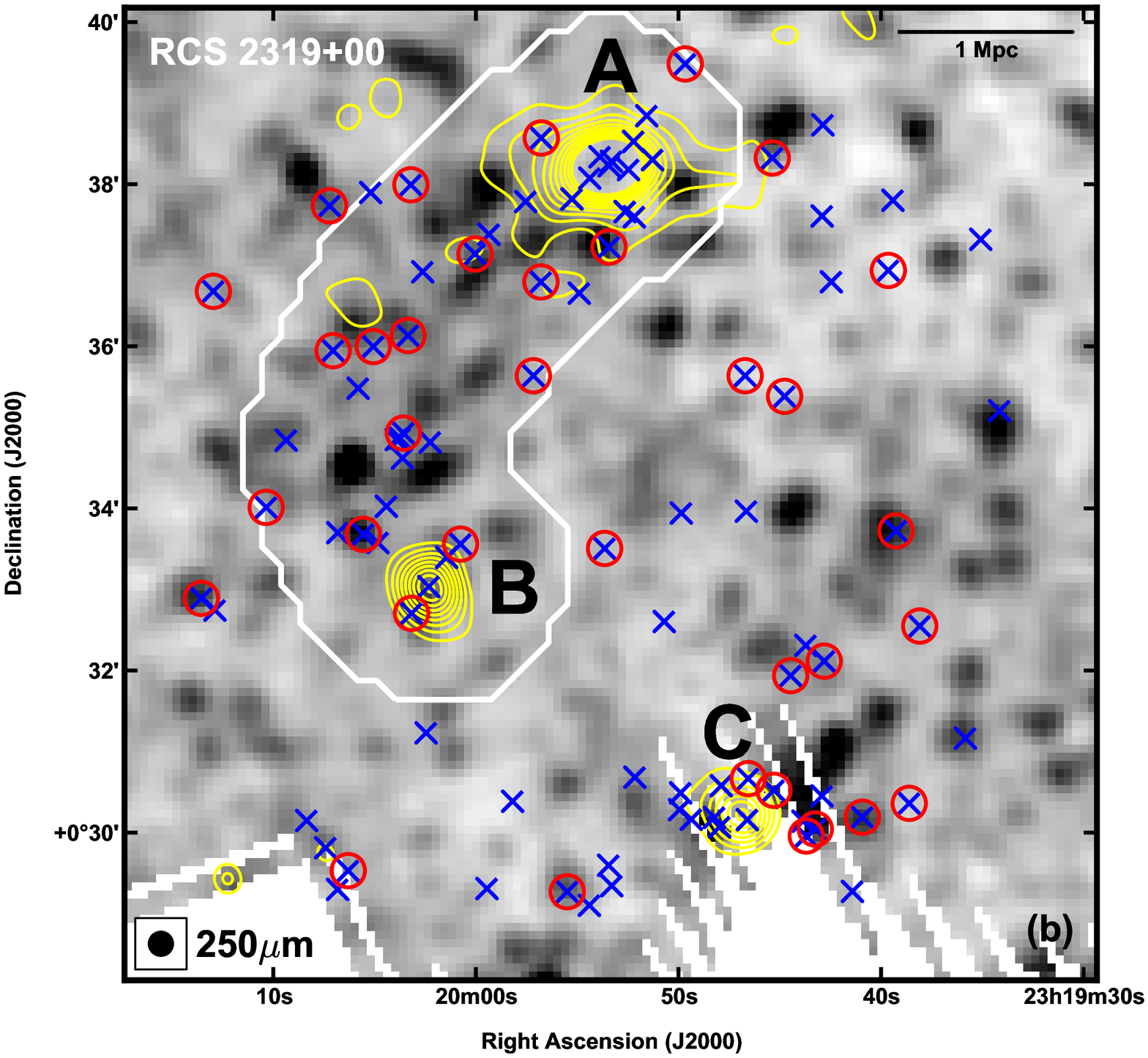}}
\centerline{\includegraphics[width=0.475\textwidth]{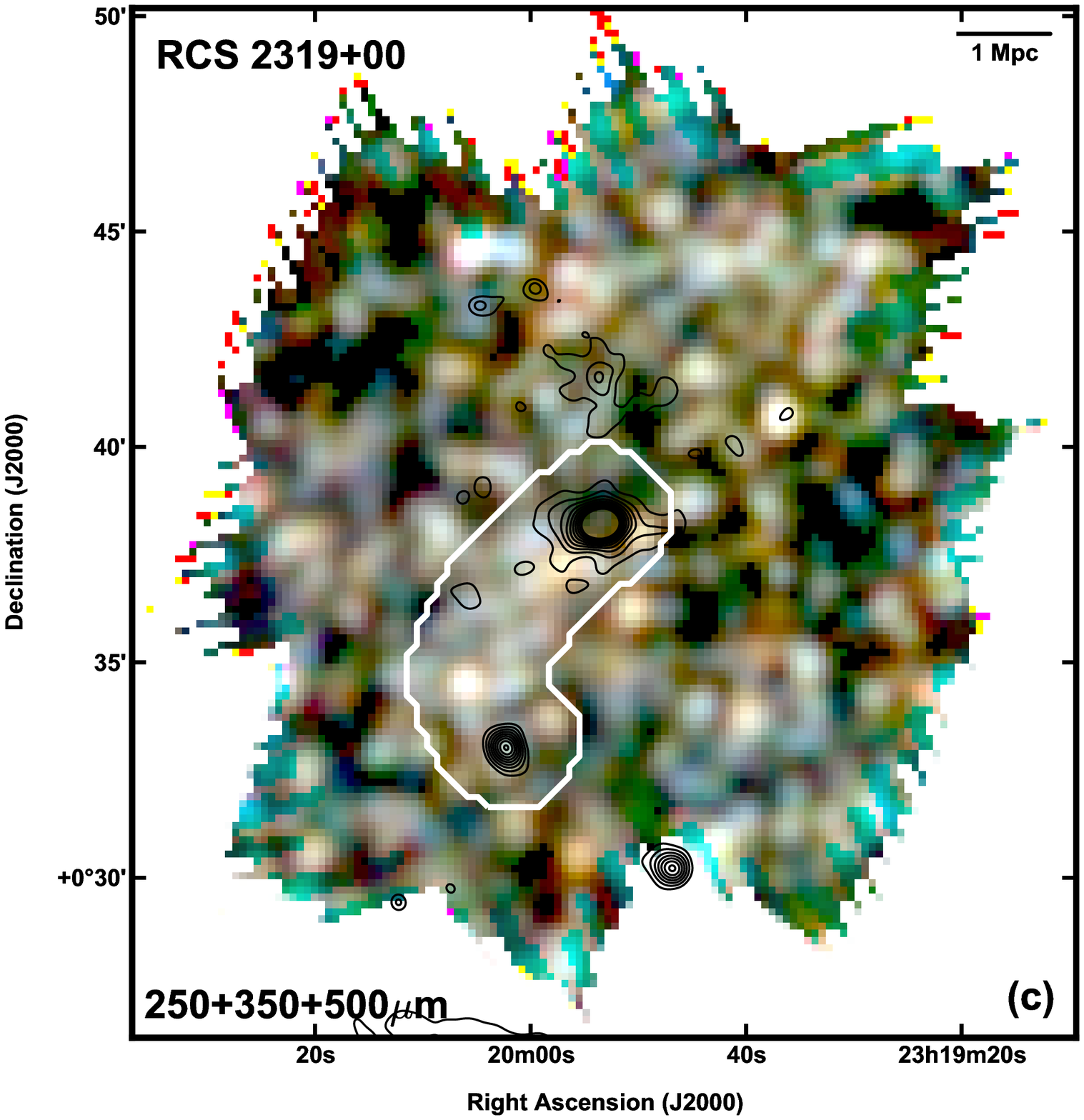}\hfill\includegraphics[width=0.475\textwidth]{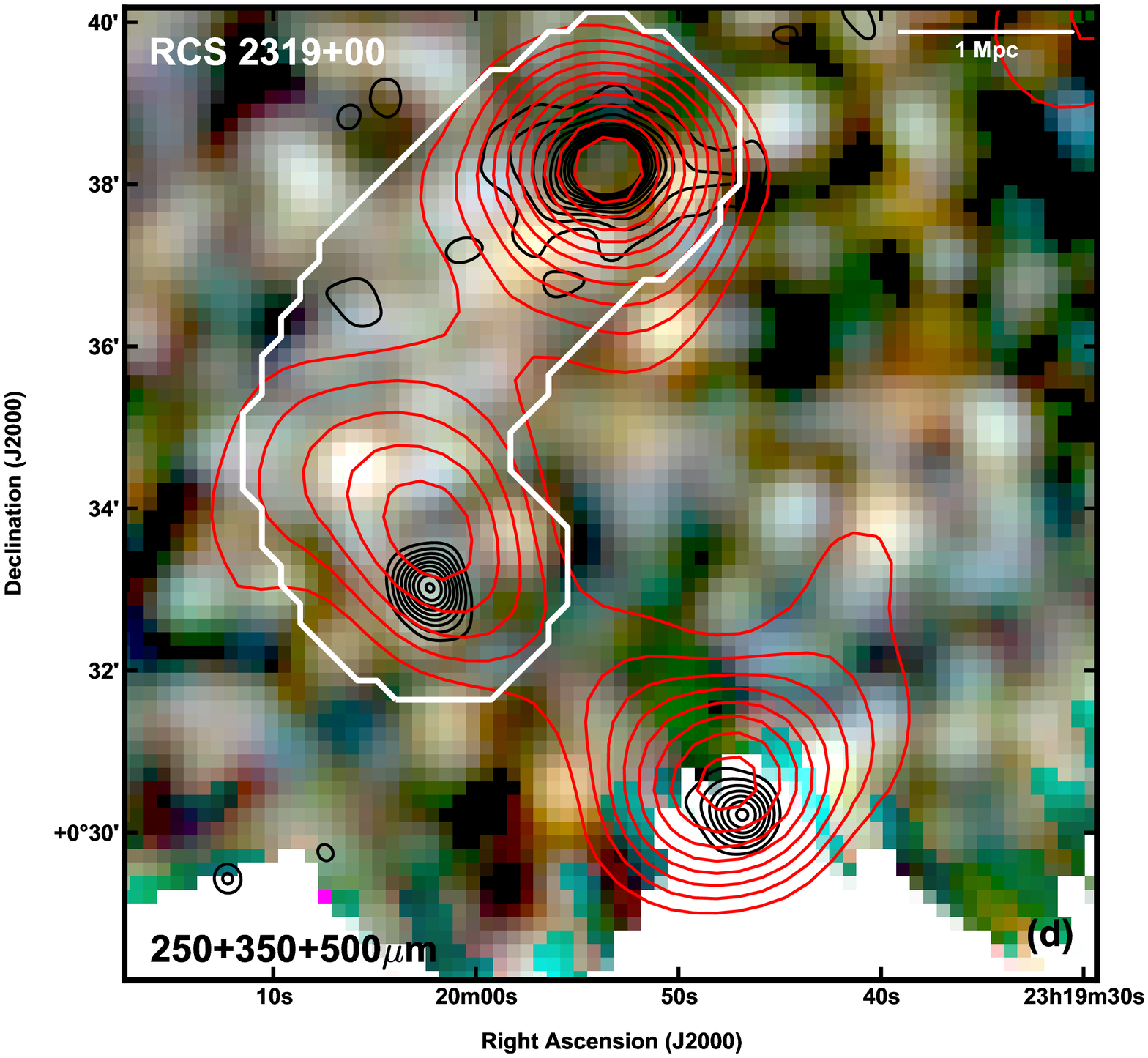}}
\caption{The {\it Herschel} filament in
  RCS\,2319+00. The white
outline indicates the filamentary region defined for the analysis, spanning
components A--B, and is overlaid on the 250\,$\mu$m SPIRE map ({\it panels
a and b}) and on an RGB composite of the 250, 350 and 500\,$\mu$m maps
smoothed to the lowest resolution ({\it panels c and d}).  Smoothed X-ray emission contours (with the point-sources
removed) are overplotted in yellow ({\it panels a and b}) or in black
({\it panels c and d}). {\it Panel b)}~~Blue `x' symbols indicate the positions of the 41 confirmed
$0.8700<z<0.9247$ cluster members, and are encircled in red if they are 24\,$\mu$m-detected.  {\it Panel d)}~~The red contours trace the surface density of galaxies
selected by the RCS selection of `red-sequence' galaxies at
$z\sim0.9$.  As pointed out in \citet{Gilbank08}, note that the major
axis of the red-sequence clusters tends to point towards its
nearest neighbour, and also notice the striking offset from the
red-sequence defined cluster cores `B' and `C' to the
peaks of the X-ray emission.}\label{fig1}
\end{figure*}

\section{Observations and data reduction}

\subsection{Herschel data}

RCS\,2319+00 was observed with \textit{Herschel} SPIRE using the
`Large Map' mode on 2009 December 17 as
part of the {\it Herschel} Lensing Survey (OD 217; OBSID 1342188181; \citealt{Egami10}). 
The scan direction was set to scan angles A and B, with the
length and height of the map set to 4$'$, yielding a map $17'\times17'$ in
extent, with a total observing time of 1.7\,hr and on-source
integration time of 0.6\,hr (17\,s per pixel). The data were
reduced with the latest version of the \textit{Herschel} interactive processing
environment ({\sc hipe} v7.3.0; \citealt{Ott10}), including the use of
the sigma-kappa deglitcher which showed improvement over the default
routine.  The maps have default pixel scales of 6, 10, and 14$''$ at 250, 350, and
500\,$\mu$m, respectively. The {\sc hipe} {\it SUSSEXtractor} task
\citep{Savage07} was used to produce
match-filtered maps (raw maps convolved with a point spread function (PSF) of 18, 25, and 36$''$ at 250, 350, and
500\,$\mu$m, respectively) and signal-to-noise ratio (SNR)
$>3$ source catalogs.  The 1\,$\sigma$ rms
instrumental noise in the match-filtered maps is $\simeq$1.1, 1.1,
and 1.7\,mJy at 250, 350, and 500\,$\mu$m, respectively. When quoting
flux uncertainties, we add in quadrature to the instrumental noise
a nominal confusion noise of $\simeq$5.8, 6.3, and 6.8\,mJy at 250,
350, and 500\,$\mu$m, respectively \citep{Nguyen10}.  The weighted SPIRE map means are consistent with 0\,mJy.

\subsection{Spitzer data}

RCS-1 imaging and photometry catalogs from the 24\,$\mu$m Multiband Imaging
Photometer for \textit{Spitzer} (MIPS; \citealt{Rieke04}) are
presented in T. Webb et al.\ (2012, in preparation).  The MIPS coverage is complete down to
$\simeq96$\,$\mu$Jy; for comparison, a local IR luminous galaxy, or LIRG, would be
$\gtrsim114$\,$\mu$Jy at the cluster redshift.  \textit{Spitzer}
InfraRed Array Camera (IRAC; \citealt{Fazio04}) four-channel imaging was
obtained through programs 30940 and 50720 and reaches M$_\mathrm{K}^\star$ + 1.5 at $z = $ 0.9 and covers approximately
the same area as the 24\,$\mu$m observations.  Image processing
was performed using {\sc IRACproc} \citep{schust06}, software
developed to wrap the existing MOsaicking and Point-source EXtraction (MOPEX) pipeline in IRAC mode and add
IRAC-specific improvements to outlier rejection.  Source detection
and aperture photometry were performed using the Picture Processing Package (PPP; \citealt{yee91}).

\section{An infrared-bright filament}\label{bridge}

All three SPIRE maps show a region of apparent `enhanced' far-IR emission
over the background -- in the form of a
complex of blended and confused point sources -- spanning the `A'
and `B' components of RCS\,2319+00, but not extending into 
the core region of component `A'.  For our analysis, we have defined a region encompassing the
filament, described by the zone within 90$''$ of a locus spanning the A--B cluster components
(see Fig.~\ref{fig1}).  
We discuss the significance of this emission below, but note that we cannot rule out
the possibility that it is the result of an overlap of the outer
regions of clusters A and B, rather than a true filament. However, we
note that no similar structure is seen between cluster cores B and C
which are similarly aligned on the sky, and the filament does not
appear to run directly between the A and B cores, but rather traces an
easterly arc (in both spectroscopy and SPIRE emission), suggestive of a real and unique structure.
In addition, the existence of the structure is supported by coincident over-densities of spectroscopic cluster members and MIPS 24\,$\mu$m number counts.

Based on extensive spectroscopy ($>$ 2000 spectra) over the
RCS\,2319+00 structure with the VLT VIMOS and Magellan IMACS
instruments (A. Faloon et al., 2012 in preparation; R. Yan et al.,
2012 in preparation), A. Faloon et
al.\ (2012, in preparation) have identified 14 24\,$\mu$m spectroscopically confirmed
cluster members in the filament with
$0.8700<z<0.9247$ and $(43\pm18)<S_{24}<(511\pm21)$\,$\mu$Jy. This
redshift range was chosen to encompass the entire supercluster
structure, based on the spectroscopic overdensity in Faloon et al.\
(2012, in preparation).
We find that the filament has a $\simeq$2$\times$ higher confirmed spectroscopic
member density compared with regions in the surrounding 10$\times$10\,arcmin$^{2}$
`field' area, with a significance of 2.5\,$\sigma$.
While our errors above do not take into account spectroscopic completeness,
we have attempted to minimize any bias in our comparison by limiting the analysis to `field' regions
with similar spectroscopic sampling (ie., $\sim$ 4
slits\,arcmin$^{-2}$) as the filament region and by also 
removing cluster members within 1\,arcmin
(0.5\,Mpc) of the cluster cores. A more detailed analysis will be presented
by A. Faloon et al.\ (2012, in preparation). There is also an apparent $\simeq3\,\sigma$ hint of an overdensity of 24\,$\mu$m
sources in the filament
(N$_{24}=(11.4\pm0.6)$\,arcmin$^{-2}$) versus the
immediate background
(N$_{24}=(9.6\pm0.2)$\,arcmin$^{-2}$).

\begin{figure}
	\centerline{\includegraphics[width=0.5\textwidth,angle=0]{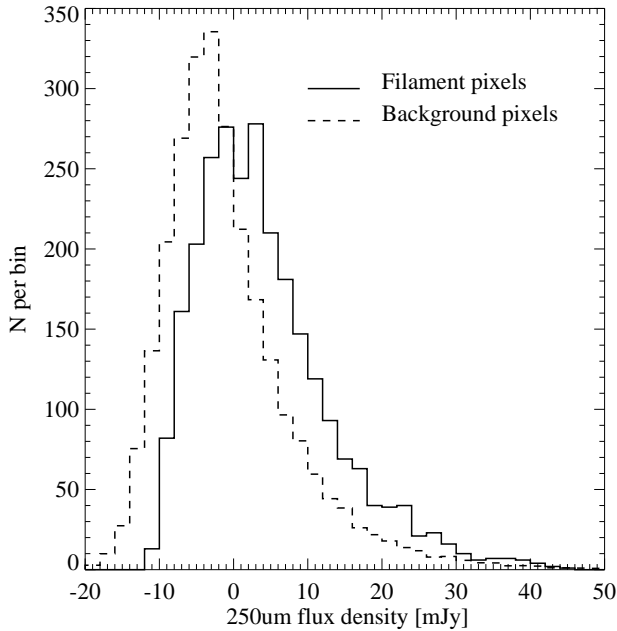}}
	\caption{Area-normalized 250\,$\mu$m pixel flux histograms showing a
          striking excess of far-IR emission in the filament
          versus the background (pixels outside the filament region).}\label{hist}
\end{figure}

To get a sense of the significance of the apparent far-IR enhancement
within the filament as revealed by {\it Herschel}, we
can simply sum the flux densities of the SNR$>3\,\sigma$ 250\,$\mu$m sources. However,
note that point source extraction is inherently insensitive to blended and confused SPIRE sources
and to extended emission, and thus our estimate will only yield a lower limit
to the true level of far-IR emission if all of the emission seen is
originating in the filament structure.  To account for
the apparent diffuse emission in the filament region,
we plot a pixel flux histogram of the filament versus the
background, showing a striking excess of far-IR emission in the
filament at all flux levels (see Fig.~\ref{hist}).  We now sum the 29 250\,$\mu$m sources in the
filament region and subtract a conservative area-normalized background estimate (based on
the flux density of SNR$>3$ sources outside of the
filament region), yielding $S_{\rm
  250}=281\pm33$\,mJy.  Due to the incompleteness in our source extraction,
unsurprisingly we see only a $\simeq2\,\sigma$ hint of an
overdensity in the filament (N$_{250}=(1.1\pm0.2)$\,arcmin$^{-2}$) versus the
background (N$_{250}=(0.7\pm0.05)$\,arcmin$^{-2}$).  

We now present a more quantitative
analysis and interpretation of the level of star-formation enhancement in the
far-IR filament using our collection of extensive follow-up
spectroscopy in RCS\,2319+00.

\subsection{Total infrared emission and star formation rate in the filament}\label{analysis}

We can estimate the contribution
of far-IR emission in the filament by summing the far-IR
flux densities of individual IR-bright cluster members and then scaling an appropriate
galaxy spectral energy distribution (SED).  To identify IR-bright cluster
members we work directly with the MIPS 24\,$\mu$m imaging which has a much
smaller PSF (FWHM$\simeq6''$) than the
SPIRE data and therefore allows for reliable cross-referencing with
the spectroscopy. 
We then determine the appropriate far-IR SED shape by performing a
standard stacking analysis (e.g.~\citealt{Marsden09};
\citealt{Pascale09}) to derive the average three-band SPIRE flux densities
of the 24\,$\mu$m-detected cluster members in the
filament, thus yielding the average SED from 24--500\,$\mu$m.  

\begin{figure}
	\centerline{\includegraphics[width=0.5\textwidth,angle=-90]{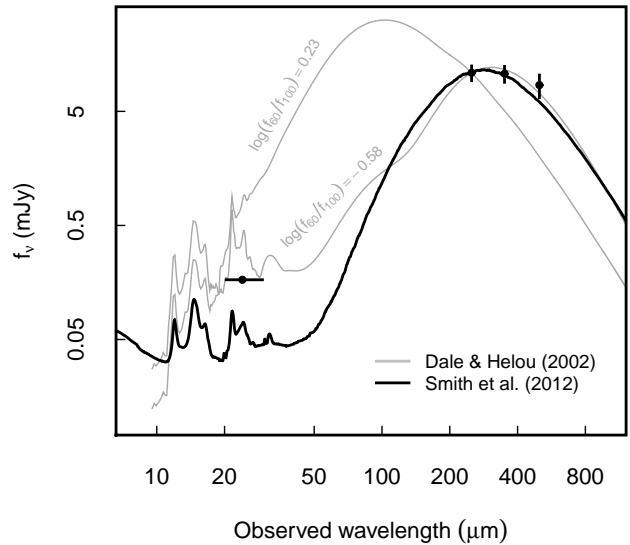}}
	\caption{SEDs redshifted to
          $z=0.9$ and normalized to the
          average (stacked) 250\,$\mu$m SPIRE flux at the 14 24\,$\mu$m cluster
          members in the filament.  Three SED templates have been
          plotted for illustrative purposes, including the two extremes of the \citet{Dale02}
          library (in terms of the $S_{60}/S_{100}$ color), as well as
          the \citet{Smith11} H-ATLAS galaxy
template with $1\times10^{11}$\,L$_{\odot}<L_\mathrm{dust}<3\times10^{11}$\,L$_{\odot}$.  The `cool' color SED templates from
          \citet{Dale02}, as well as the H-ATLAS SED are well matched to the average far-IR
emission of MIPS-selected emitters in the filament. At $z=0.9$, the 24\,$\mu$m band is tracing the rest-frame 13\,$\mu$m
light, which is dominated by a complex of broad PAH emission, thus we expect some
variation in SED template matches in this band, as is seen here.}\label{fig2}
\end{figure}

The SPIRE imaging is much shallower relative to the MIPS imaging, and
is only sensitive to L$_{\rm IR}\gtrsim10^{12}$\,L$_\odot$ activity at
the cluster redshift -- and indeed only three of the 24\,$\mu$m cluster members
appear to be `detected' at 250\,$\mu$m (see Fig.~\ref{fig1}).  We thus proceed
to stack the flux densities from the SPIRE maps at the location
of the 14 MIPS cluster members (which have an average
24\,$\mu$m flux of $\simeq170$\,$\mu$Jy), yielding significant stacked
flux densities of $11.0\pm1.6$\,mJy (7.0\,$\sigma$), $10.7\pm1.7$\,mJy
(6.2\,$\sigma$), and $8.5\pm1.9$\,mJy (4.6\,$\sigma$) at 250, 350, and
500\,$\mu$m, respectively.   We note that stacking the cluster members
on the SPIRE maps using the much more precise MIPS positions should
help to minimize any blending and confusion contamination from the
SPIRE emission of non-cluster members. 
Fig.~\ref{fig2} shows the stacked SPIRE flux densities in
relation to a set of spectral templates \citep{Dale02} and the median {\it Herschel}-Astrophysical Terahertz Large
Area Survey (H-ATLAS; \citealt{Eales10}) galaxy
template with $1\times10^{11}$\,L$_{\odot}<$L$_\mathrm{dust}<3\times10^{11}$\,L$_{\odot}$
from \citet{Smith11}, redshifted to $z=0.9$ and normalized to the
250\,$\mu$m stacked flux density.  The far-IR colors are consistent with
the `cool' templates from the
\citet{Dale02} library, with $S_{60}/S_{100}\lesssim0.26$ rest-frame colors. 
The H-ATLAS template is also a reasonable fit
to the average far-IR emission, yielding
L$_{\rm IR}=(1.9\pm0.3)\times10^{11}$\,L$_{\odot}$, which is
consistent with the LIRG-class.  Using the \citet{Kennicutt98} SFR conversion, SFR\,(M$_{\odot}$\,yr$^{-1}$)$=4.5\times10^{-44}($L$_{\rm
  IR}/{\rm erg\,s^{-1}})$ (which assumes a starburst age $<$100\,Myr 
and a \citealt{Salpeter55} initial mass function), yields an average SFR of $(30\pm5)$\,M$_{\odot}$\,yr$^{-1}$.

At $z=0.9$, the MIPS 24\,$\mu$m band is dominated by a complex of broad
polycyclic aromatic hydrocarbon (PAH) emission. When assessing the
total L$_\mathrm{IR}$, as defined by the 8--1000\,$\mu$m integrated emission,
normalizing template SEDs to 24\,$\mu$m emission alone is subject to a high degree of uncertainty, due to the
potential variation (intrinsic and redshifted) of the PAH contribution in the 24\,$\mu$m band. Similarly, our lack of spectral coverage in the
30--100\,$\mu$m range means we have poor constraints on the
contribution of a hot dust continuum component, which could boost the
luminosity in the 24\,$\mu$m band (e.g.,~Fig.~\ref{fig2}). This is evidenced by the range of 24\,$\mu$m fluxes
predicted by the 250\,$\mu$m normalized templates, which straddle the average
24\,$\mu$m emission. Therefore, we take the \citet{Smith11} template as a
conservative estimate of the form of the RCS\,2319+00 members' SEDs, and use this
to calculate the total L$_\mathrm{IR}$ of the filament.

Since our spectroscopic follow-up of the MIPS cluster members is
incomplete, we can only place a conservative lower limit to the total far-IR emission in the filament by summing the 250\,$\mu$m
emission contributed by the 14 24\,$\mu$m-detected cluster members, yielding $S_{\rm 250}=154\pm22$\,mJy.  
Assuming each galaxy has the same SED shape, we
use the H-ATLAS template to determine a
total L$_{\rm IR}=(2.6\pm0.4)\times10^{12}$\,L$_\odot$, corresponding to a total SFR within the
filament of $(460\pm70)$\,M$_\odot$\,yr$^{-1}$.  
We stress that these estimates are strictly lower limits due to the spectroscopic
incompleteness of the galaxies in the vicinity of the filament and the requirement of a 24$\mu$m detection. For example, if we include the
27 cluster members that were not detected at 24\,$\mu$m in the stack, we find
a factor of 40\% more far-IR emission.  

Rather than attempt to correct for spectroscopic completeness we now
use a different approach to estimate the total IR emission. Earlier we estimated the statistical total IR excess
emission within the filament compared to the average over the
map using only the SPIRE data.  This should provide a more complete picture of the level of star-formation activity in the
filament as this estimate does not suffer from spectroscopic
incompleteness.  The total (background subtracted) signal is $S_{\rm 250}=281\pm33$\,mJy within
an area of 26.3\,arcmin$^{2}$, which corresponds to a total L$_{\rm
IR}=(4.8\pm0.6)\times10^{12}$L$_\odot$ and SFR within the filament of $\sim 850\pm100$\,M$_\odot$\,yr$^{-1}$.

\section{Discussion}\label{discuss}

How can we interpret these observations? First we note that star formation in
the filament galaxies is unlikely to be impeded by ram-pressure
stripping, given (a) the relatively low density of the intergalactic
medium (IGM; \citealt{Dave01}) and
relative galaxy velocities compared to the cores of the clusters (ram-pressure
scales with the ICM density and the square of the galaxy velocity;
\citealt{Gunn72}); and (b) the relatively short duration for the star
formation events (100s\,Myr) compared to the time required for stripping in
these environments (up to several Gyr, e.g.,~\citealt{McCarthy08}).
Furthermore, conditions within intermediate density (i.e.,~filament)
environments might be suitable for tidally inducing activity via close
interactions and mergers between gas-rich systems being accreted onto such
environments. For example, \citet{Geach11} find an enhancement in the local
fraction of (UV/IR selected) star-forming galaxies in a similar
filamentary environment surrounding the Cl\,0016+16 supercluster at $z=0.55$.
Is the star formation in the RCS\,2319+00 supercluster filament significantly
enhanced compared to the field?

We have established that the total SFR occurring within the
filament, as traced by the 250\,$\mu$m emission, is significantly larger than the
local background.  On its own, however, this does not directly indicate
triggered or enhanced star formation as this may simply reflect the underlying
over-density of star-forming galaxies which have not been actively quenched. 
To quantify any possible enhancement we look at the SFR per unit stellar
mass, M$_\star$, or specific SFRs (SSFRs) of the MIPS cluster
members in the filament. The stellar mass was derived directly from the
measured IRAC-4.5\,$\mu$m 
flux densities after applying an $(R_c-z^\prime)$ color correction
to the M/L ratio, following \citet{Bell03}.  Normalized to the same
stellar mass, the individual IR galaxies within the filament have
$\sim$3$\times$ the SSFR of IR-bright galaxies within the central
cluster cores, but only $\sim1.5\times$ that of galaxies in the rest
of the supercluster.  To avoid systematic differences we do not
attempt to compare this to field galaxies at this epoch and simply note that the filament
galaxies do not show enhanced emission compared to the rest of the supercluster. 

How important would this star formation be to the overall growth of what will be a very massive cluster?
Given the apparent low stellar mass of the individual galaxies within the filament, their eventual
fate is to fall onto the faint-end of the red sequence.  We have
estimated the total stellar mass already in place
in RCS\,2319+00--A within 2.1\,Mpc to be M$_\star \sim
3\times10^{12}$\,M$_\odot$.  
This estimate is obtained by
fitting a Schechter function to the background-subtracted 4.5\,$\mu$m
flux and integrating this to provide a total rest-frame $K$-band luminosity (above
M$_\mathrm{AB}\sim -22$) which is then converted to stellar mass assuming an early-type SED \citep{sb99}.  
Using the M$_{\rm tot}$ derived from the X-ray luminosity provides a stellar-to-total
mass conversion which we then apply to the other two cluster cores,
yielding a total M$_\star$($<$2.1\,Mpc)$\sim7\times10^{12}$\,M$_\odot$. Thus, if the filament galaxies continue at their current SFRs
until the clusters finally merge at $z\sim$0.5,  they will increase
the descendent cluster mass by 20\%.   This should be regarded as an
upper-limit, as the star formation will begin to shut down at earlier
times as the galaxies enter the increasingly dense cluster environment and
as fuel is exhausted; nevertheless the pre-processing witnessed in this filament appears to be
an important stage of the build-up of the cluster stellar mass. 

\section*{Acknowledgements}

We thank an anonymous referee for a helpful report which improved the
paper.  We are grateful to Dan Smith for allowing the pre-publication use of the H-ATLAS SED.  
KEKC acknowledges support from the endowment of the Lorne Trottier
Chair in Astrophysics and Cosmology at McGill, the Natural Science and Engineering
Research Council of Canada (NSERC), and a L'Or\'{e}al Canada Women in Science
Research Excellence Fellowship.   JEG is supported by a Banting Postdoctoral
Fellowship and AF is funded through an NSERC PGSD, both administered
by NSERC.  TMAW acknowledges support from NSERC. The {\it Herschel}
spacecraft was designed, built, tested, and launched under a contract
to ESA managed by the Herschel/Planck Project team by an industrial
consortium.
This work is based in part on observations made with the
\textit{Spitzer Space Telescope}, which is operated by JPL, Caltech under a
contract with NASA.

\setlength{\bibhang}{2.0em}

\label{lastpage}

\begin{thebibliography}{50}
\setlength{\itemindent}{-2.5em}
\bibitem[Bai et al.(2009)]{Bai09}Bai, L., Rieke, G.H.,
Rieke, M.J., et al., 2009, ApJ, 693, 1840
\bibitem[Bell et al.(2003)]{Bell03}Bell, E.F., McIntosh, D.H., Katz,
  N., Weinberg, M.D., 2003, ApJS, 149, 289
\bibitem[Butcher \& Oemler(1978)]{BO78}Butcher, H., \& Oemler, A., Jr.,
  1978, ApJ, 226, 559
\bibitem[Butcher \& Oemler(1984)]{BO84}Butcher, H., \& Oemler, A., Jr.,
  1984, ApJ, 285, 426
\bibitem[Dale \& Helou(2002)]{Dale02}Dale, D.A., \& Helou, G., 2002, ApJ,
  576, 159
\bibitem[Dav\'{e} et al.(2001)]{Dave01}Dav\'{e}, R., Cen, R., Ostriker,
  J.P., et al., 2001, ApJ, 552, 473
\bibitem[De Lucia et al.(2004)]{DeLucia04}De Lucia, G., Poggianti,
  B.M., Arag\'{o}n-Salamanca, A., et al., 2004, ApJ, 610, L77
\bibitem[Eales et al.(2010)]{Eales10}Eales, S., Dunne, L., Clements, D., et al., 2010, PASP, 122,
  499
\bibitem[Egami et al.(2010)]{Egami10}Egami, E., Rex, M., Rawle, T.D., et al., 2010, A\&A, 518,
  L12
\bibitem[Fazio et al.(2004)]{Fazio04}Fazio, G.G., Ashby, M.L.N.,
  Barmby, P., et al., 2004, ApJS, 154, 10
\bibitem[Geach et al.(2006)]{Geach06}Geach, J.E., Smail, I., Ellis,
  R.S., et al., 2006, ApJ, 649, 661
\bibitem[Geach et al.(2011)]{Geach11}Geach, J.E., Ellis, R.S., Smail,
  I., Rawle, T.D., Moran S.M., 2011, MNRAS, 413, 177 
\bibitem[Gilbank \& Balogh(2008)]{GilBal08}Gilbank, D.G., \& Balogh,
  M.L., 2008, MNRAS, 385, L116
\bibitem[Gilbank et al.(2008a)]{Gilbank08LF}Gilbank, D.G., Yee,
  H.K.C., Ellingson, E., et al., 2008a, ApJ, 673, 742
\bibitem[Gilbank et al.(2008b)]{Gilbank08}Gilbank, D.G., Yee, H.K.C.,
  Ellingson, E., et al., 2008b, ApJ, 677, L89 
\bibitem[Gladders \& Yee(2005)]{GladdersYee05}Gladders, M.D., \&
  Yee, H.K.C., 2005, ApJS, 157, 1
\bibitem[Griffin et al.(2010)]{Griffin10}Griffin, M.J., Abergel, A.,
  Abreu, A., et al. 2010, A\&A, 518, L3
\bibitem[Gunn \& Gott(1972)]{Gunn72}Gunn, J.E., \& Gott, J.R., III, 1972, ApJ,
  176, 1
\bibitem[Hicks et al.(2008)]{Hicks08}Hicks, A.K., Ellingson, E.,
  Bautz, M., et al., 2008, ApJ, 680, 1022
\bibitem[Hopkins \& Beacom(2006)]{Hopkins06}Hopkins, A.M., \& Beacom, J.F.,
  2006, ApJ, 651, 142
\bibitem[Kennicutt(1998)]{Kennicutt98}Kennicutt, R.C., 1998, ARA\&A,
  36, 189
\bibitem[Leitherer et al.(1999)]{sb99}Leitherer, C., Schaerer, D.,
  Goldader, J.D., et al., 1999, ApJS, 123, 3
\bibitem[Marsden et al.(2009)]{Marsden09}Marsden, G., Ade, P.A.R., Bock,
  J.J., et al., 2009, ApJ, 707, 1729
\bibitem[McCarthy et al.(2008)]{McCarthy08}McCarthy, I.G., Frenk,
  C.S., Font A.S., et al., 2008, 383, 593
\bibitem[Nguyen et al.(2010)]{Nguyen10}Nguyen, H.T., Schulz, B.,
  Levenson, L., et al., 2010, A\&A, 518, L5
\bibitem[Ott(2010)]{Ott10}Ott, S., 2010, ASP Conference Series, 434,
  139
\bibitem[Papovich et al.(2010)]{Papovich10}Papovich, C., Momcheva, I.,
  Willmer, C.N.A., et al., 2010, ApJ, 716, 1503
\bibitem[Pascale et al.(2009)]{Pascale09}Pascale, E., Ade, P.A.R., Bock,
  J.J., et al., 2009, ApJ, 707, 1740
\bibitem[Pilbratt et al.(2010)]{Pilbratt10}Pilbratt, G.L., Riedinger,
  J.R., Passvogel, T., et al., 2010, A\&A, 518, L1
\bibitem[Poggianti et al.(2008)]{Poggianti08}Poggianti, B.M., Desai,
  V., Finn, R., et al., 2008, ApJ, 684, 888
\bibitem[Rieke et al.(2004)]{Rieke04}Rieke, G., Young, E.T.,
  Engelbracht, C.W., et al., 2004, ApJS, 154, 25
\bibitem[Salpeter(1955)]{Salpeter55}Salpeter, E.E., 1955, ApJ, 121, 161
\bibitem[Savage \& Oliver(2007)]{Savage07}Savage, R.S., \& Oliver, S.,
  2007, ApJ, 661, 1339
\bibitem[Schuster, Marengo \& Patten(2006)]{schust06}Schuster, M.T.,
  Marengo, M., Patten, B.M., 2006, SPIE, 6270, 65
\bibitem[Smith et al.(2012)]{Smith11}Smith, D.J.B., Dunne, L., da Cunha,
  E., et al., 2012, MNRAS, submitted
\bibitem[Spergel et al.(2003)]{Spergel03}Spergel, D.N., Verde, L.,
  Peiris, H.V., et al., 2003, ApJS, 148, 175
\bibitem[Yee(1991)]{yee91}Yee, H.K.C., 1991, PASP, 103, 396

\end{thebibliography}
\end{document}